\documentclass[twocolumn,aps,10pt,prl,showpacs,showkeys,]{revtex4}
\usepackage[dvips]{graphicx}

\begin{document}
\title{Comment on a Phys. Rev. Lett. paper: the origin of the excited state in LaCoO$_{3}$}
\author{R. J. Radwanski}
\affiliation{Center of Solid State Physics, S$^{nt}$Filip 5,
31-150 Krakow, Poland \\
Institute of Physics, Pedagogical University, 30-084 Krakow,
Poland}\homepage{http://www.css-physics.edu.pl}
\email{sfradwan@cyf-kr.edu.pl}
\author{Z. Ropka}
\affiliation{Center of Solid State Physics, S$^{nt}$Filip 5,
31-150 Krakow, Poland}

\begin{abstract}
In contrary to a claim of the recent Phys. Rev. Lett. {\bf 96}
(2006) 027201 paper we maintain that the first excited state in
LaCoO$_{3}$ is the high-spin (HS) state (a lowest quasi-triplet
from the octahedral subterm $^{5}T_{2g}$ of the $^{5}D$ term,
Phys. Rev. B {\bf 67} (2003) 172401) in agreement with the
Tanabe-Sugano diagram.

\vspace {-1.6 cm}
\end{abstract}
\maketitle \vspace {-1.6 cm} In a recent paper Phelan {\em et~al.}
\cite {1} claim that the excited state in LaCoO$_{3}$ is the
intermediate-spin (IS) $S$=1 state of a $t_{2g}^{5}e_{g}^{1}$
configuration.

By this Comment we would like to correct this claim. We would
expect that the problem of the excited state in LaCoO$_{3}$ has
been clarified in a year of 2003 in our paper \cite {2}, making
use of experimental results of Noguchi {\em et~al.} \cite{3}, but
authors of the commented paper likely did not notice this paper.
They have cited our first paper about LaCoO$_{3}$ from a year of
1999 in Ref. 22 - in fact this paper dealt with the splitting of
of the $^{5}D$ term, belonging to the high-spin
$t_{2g}^{4}e_{g}^{2}$ ($S$=2) state, by the trigonal distortion
in the presence of the spin-orbit coupling.

In Ref. \cite{2} we have proved that the excited state
(quasi-triplet) originates from the $^{5}T_{2g}$ subterm of the
$^{5}D$ term belonging to the HS $t_{2g}^{4}e_{g}^{2}$ ($S$=2)
configuration.  We have perfectly reproduced the magnetic-field
behaviour of the quasi-triplet and the anisotropic $g$ factor
experimentally revealed by Noguchi {\em et~al.} \cite{3}. The
ground state is a many-electron subterm $^{1}A_{1}$ originating
from the $^{1}I$ term, which in the free Co$^{3+}$ ion lies 4.45
eV above the ground term $^{5}D$. The 13-fold degenerated $^{1}I$
term is split by the octahedral crystal field and the subterm
$^{1}A_{1}$ is strongly pushed down, by relatively strong crystal
field, due to its very large orbital quantum number $L$=6, as
occurs on the Tanabe-Sugano diagram for $Dq$/$B$=2.025.

The IS state as the first excited state has been introduced to
the LaCoO$_{3}$ problem in year of 1996 by band calculations of
Korotin {\em et~al.} \cite{4} as an {\bf opposite} view to the
atomistic view being a base for the Tanabe-Sugano diagrams known
from years of 1954. The Tanabe-Sugano diagram for the 3$d^{6}$
configuration has yielded the excited state to be the HS state
and this view was the base for a model of Goodenough. The
IS-state concept has become highly popular \cite {5,6}. In the
band calculations of Korotin {\em et~al.} the IS state becomes the
first excited state as an effect of the especially strong $d-p$
hybridization. However, we claim that if at present, in a year of
2006, one wants to still claim that the IS state is an excited
state has to present a quantitative band-based or
hybridization-based interpretation of the Noguchi {\em et~al.}
experiment. In the atomic physics $t_{2g}^{5}e_{g}^{1}$ ($S$=1)
state is 24-fold degenerated - thus there is a question about a
degeneracy left in LaCoO$_{3}$ and its characteristics.

In the ionic atomistic (QUASST) picture the discrete atomiclike
electronic structure is preserved also in transition-metal solid.
For instance, the meV-scale splitting of the 15-fold degenerated
$^{5}T_{2g}$ subterm by trigonal distortion in the presence of
the spin-orbit coupling has been presented in Ref. \cite{2} for
LaCoO$_{3}$ and for the Fe$^{2+}$ ion in FeBr$_{2}$. We note that
the meaning and the degeneracy of the LS, IS and HS states in the
band picture is understood differently than in the ionic (QUASST)
picture. Thus, we think that the basic problem of LaCoO$_{3}$ is
associated with a consideration of $d$ states as localized
(ionic, QUASST) or as delocalized forming a wide energy $\sim$10
eV band like in Ref. \cite 4 and to settle down the $d$
occupation/valency of 6/+3 or 7.3/+1.7. Within the localized
picture the estimation of the strength of the octahedral
crystal-field interactions is decisive.\vspace {-0.2 cm}

{\bf In conclusion}, we claim that the origin of the excited state
in LaCoO$_{3}$ has been already established to be the high-spin
state in agreement with the Tanabe-Sugano diagram. 

\end{document}